\documentclass[conference]{IEEEtran}

\usepackage{cite}
\usepackage{amsmath,amssymb,amsfonts}
\usepackage{graphicx}
\usepackage{textcomp}
\usepackage{xcolor}
\usepackage{float}
\usepackage{booktabs}

\usepackage{algorithm}
\usepackage{algpseudocode}

\def\BibTeX{{\rm B\kern-.05em{\sc i\kern-.025em b}\kern-.08em
    T\kern-.1667em\lower.7ex\hbox{E}\kern-.125emX}}

\usepackage{background}
\usepackage{xcolor}
\usepackage{hyperref}

\backgroundsetup{
  scale=1,
  color=black,
  opacity=1,
  angle=0,
  position=current page.south,
  vshift=10pt,
  contents={\textcolor{red}{© 2026. For personal use only. Published version under doi:  \href{https://doi.org/10.1109/WiMob70978.2026.11777854}{10.1109/WiMob70978.2026.11777854}. }}
}

\begin{document}

\title{A Reactive Cross-Layer QoS Adaptation Framework for SDN-Enabled DDS Communications in IoD}

\author{\IEEEauthorblockN{Mohamed Ali Zormati}
\IEEEauthorblockA{\textit{Université de Technologie de Compiègne} \\
\textit{CNRS, Heudiasyc UMR 7253}\\
Compiègne, France \\
mohamed-ali.zormati@hds.utc.fr}
\and
\IEEEauthorblockN{Ghada Jaber}
\IEEEauthorblockA{\textit{Université de Technologie de Compiègne} \\
\textit{CNRS, Heudiasyc UMR 7253}\\
Compiègne, France \\
ghada.jaber@hds.utc.fr}
\and
\IEEEauthorblockN{Hicham Lakhlef}
\IEEEauthorblockA{\textit{Université de Bordeaux, CNRS} \\
\textit{Bordeaux INP, LaBRI UMR 5800}\\
Talence, France \\
hicham.lakhlef@labri.fr}
}

\maketitle

\begin{abstract}
The Internet of Drones (IoD) is a subset of the Internet of Things (IoT) where drones operate as networked smart objects communicating with entities across terrestrial, aerial, and spatial domains. The increasing deployment of drones has accelerated the adoption of IoD in both mainstream and emerging applications. Increasingly, IoD systems rely on the standardized Data Distribution Service (DDS) as their communication middleware, thanks to its rich Quality of Service (QoS) policies, which enable meeting the demanding constraints of IoD applications. However, QoS policies, particularly reliability modes, are often statically configured, which limits their effectiveness in IoD environments that are characterized by high mobility, frequent network impairments, and dynamic topology changes. In this paper, we propose a reactive QoS adaptation framework for DDS-based IoD systems. The framework leverages Software Defined Networking (SDN) to capture global network conditions and incorporates device energy levels and topic popularity to dynamically adapt reliability modes to changing operating environments. By jointly considering network-level, device-level, and topic-level inputs, the proposed approach enhances middleware adaptability, reliability, and efficiency. Simulation results demonstrate the effectiveness of the proposed framework compared to static QoS configurations.
\end{abstract}


\section{Introduction}
The Internet of Drones (IoD) is a rapidly evolving paradigm that extends the Internet of Things (IoT) into the aerial domain \cite{boccadoroExtensiveSurveyInternet2021}. The IoD interconnects ground, aerial, and space networks, thereby transforming Unmanned Aerial Vehicles (UAVs) from isolated systems into aerial agents capable of intelligent sensing, computing, and communications. This enables a wide range of applications that extend beyond terrestrial IoT \cite{abualigahApplicationsDeploymentsIntegration2021}. IoD spans from mainstream applications, such as emergency response and telecommunications infrastructure support, to enabling other emerging technologies, such as Space-Air-Ground Integrated Networks (SAGINs). Given that the effectiveness of IoD largely depends on its communication capabilities, it is crucial to develop reliable and efficient communication frameworks that can support IoD operations amid its continued growth \cite{bineConnectingInternetDrones2024}. This objective is challenged by the inherent constraints of drones, such as high mobility and limited resources, as well as the instability of wireless links \cite{javedStateoftheArtFutureResearch2024}.

Increasingly, IoD systems rely on the standardized Data Distribution Service (DDS) as their communication middleware \cite{leeOptimizingDataDistribution2025} thanks to its data-centric publish/subscribe model design, rich QoS policies, and distributed architecture \cite{gamboSystematicLiteratureReview2025}. Several recent works have investigated the use of DDS in IoD, highlighting its suitability for real-time coordination and scalable data dissemination \cite{vidalEnablingMultiMissionInteroperable2018}. However, most existing works rely on static QoS configurations, which are challenged by the highly dynamic conditions of IoD environments \cite{ingles-romeroModelDrivenApproachEnable2017}. Despite the rich QoS support provided by DDS, static configurations fail to respond to network variability and impairments \cite{yoonQoSOptimizer2016}. Therefore, dynamic QoS adaptation, leveraging network-level state, becomes essential.

To enable network-level input for dynamic QoS adaptation, it is relevant to consider network softwarization techniques, mainly Software Defined Networking (SDN). SDN provides global visibility of the underlying IoD network and enables dynamic, network-wide control \cite{zormatiReviewAnalysisRecent2024}. The combination of SDN with DDS has been explored to improve the latter real-time performance \cite{choiMakingDDSReally2016} and to automate deterministic deployments of DDS over Time-Sensitive Networking (TSN) \cite{liEnablingDeterministicTransmission2025}. The combination of DDS and SDN enables the flexibility required for constrained networks \cite{bertauxDDSSDNBased2014a}, such as IoD.

While several efforts have been made to enhance DDS to meet the stringent requirements of IoD, to the best of our knowledge, no prior work has proposed an SDN-based cross-layer QoS adaptation framework for DDS in the IoD context. To address this gap, we propose a novel framework that dynamically adjusts DDS QoS configurations based on network-level, device-level, and topic-level inputs. Unlike generic adaptive frameworks, the proposed solution accounts for global network conditions through SDN-based monitoring, as well as device energy levels and topic popularity. Based on these inputs, the framework dynamically adapts DDS reliability modes to enhance communication reliability and efficiency.

The remainder of the paper is organized as follows. Section II reviews related work. Section III presents the system model, and Section IV describes the proposed adaptation solution. Section V presents the experimental results. Section VI concludes the paper.

\section{Related Work}

DDS, standardized by the Object Management Group (OMG) \cite{OMG_DDS_1_4}, has been increasingly adopted in IoD systems thanks to its flexible data-centric publish/subscribe architecture and comprehensive QoS support \cite{gamboSystematicLiteratureReview2025}. DDS provides a set of 22 QoS policies that enable fine-grained control over multiple aspects of data dissemination, including reliability, latency, durability, and data availability \cite{zhangComparisonMiddlewaresEdgetoEdge2024}. These policies are configured at the topic, publisher/subscriber, and data writer/reader levels. Communication is only established when the corresponding QoS settings are compatible.

Several works have investigated runtime QoS adaptation in DDS-based systems to improve behavior under challenging operating conditions. In \cite{yoonQoSOptimizer2016}, the authors propose a monitoring-driven QoS optimizer that observes DDS runtime events and iteratively adjusts QoS profiles. In \cite{ingles-romeroModelDrivenApproachEnable2017}, a model-driven framework is introduced to enable safe runtime adaptation of DDS QoS through constrained templates, ensuring compatibility among participants while adapting to fluctuating resources. Authors in \cite{alaerjanFormalizingSemanticsDDS2023} formalize the semantics and interdependencies of DDS QoS policies. While these works demonstrate that DDS QoS parameters can be adapted at runtime, they rely on middleware- or configuration-level mechanisms and local observations, without incorporating global network context or cross-layer inputs.

To enable network-aware QoS enforcement, several efforts have explored the integration of DDS with SDN. In \cite{bertauxDDSSDNBased2014a}, the authors propose a combined architecture in which DDS communicates application-level QoS requirements to an SDN controller, which then configures network resources accordingly. In \cite{choiMakingDDSReally2016}, an SDN-based solution is introduced to enforce DDS QoS requirements at the network layer. Authors in \cite{liEnablingDeterministicTransmission2025} further extend this integration to deterministic TSN environments. While these studies highlight the benefits of network programmability, they treat DDS QoS configurations as static inputs where QoS policies are enforced by the network rather than dynamically adapted at the middleware level.

In constrained environments, energy-aware communication is essential. In \cite{al-roubaieyEATDDSEnergyawareMiddleware2019}, the authors propose an energy-aware DDS middleware that incorporates local energy models to reduce communication overhead. In parallel, other works, such as \cite{salehiPopSubImprovingResource2017}, show that data dissemination cost is strongly influenced by topic popularity. While these works highlight the importance of energy efficiency and popularity awareness, they consider these factors independently.

To the best of our knowledge, no prior work has proposed a runtime DDS reliability adaptation approach that jointly considers global network awareness via SDN, device-level energy constraints, and topic popularity. This work addresses this gap by introducing a cross-layer, SDN-assisted DDS reliability adaptation framework that dynamically adjusts DDS QoS reliability configuration based on network-level, device-level, and topic-level inputs, with the objective of enhancing communication reliability and efficiency.

\section{System Model}
We consider a DDS-based IoD network, composed of a set of drones $D$. We denote by $l_{ij}$ the wireless link between drones $i$ and $j$, with $i,j \in D$, and $l_{ij} \in L$, where $L$ is the set of available wireless links. Such an IoD network is subject to both energy constraints, due to battery-powered drones, and communication reliability challenges arising from unreliable wireless links.

To ensure mission success, drones exchange information through distinct topics $\tau \in T$, where $T$ denotes the set of topics. Each topic is associated with specific QoS requirements. For clarity, we focus on DDS reliability modes, which can be configured as $\left\{ \texttt{BEST\_EFFORT},\; \texttt{RELIABLE} \right\}$ \cite{OMG_DDS_1_4}. The reliable mode ensures delivery via acknowledgments and retransmissions at the cost of higher latency and resource usage, whereas the best effort mode provides no guarantees but favors low latency and overhead.

\begin{figure}[ht]
    \centering
    \includegraphics[scale=0.2]{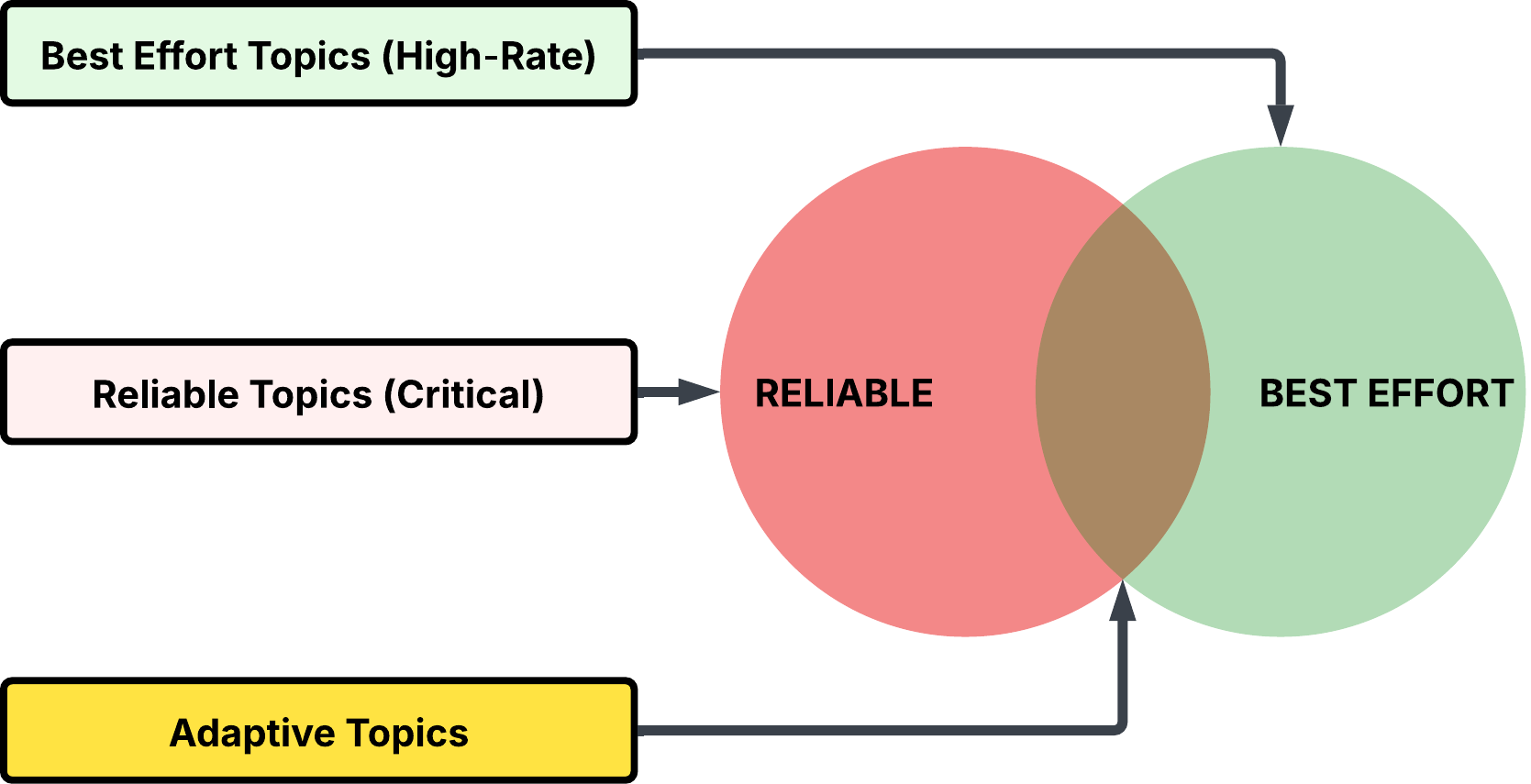}
    \caption{Reliability-based topic categorization}
    \label{fig_rel}
\end{figure}

For certain critical topics, such as safety and control information (e.g., emergency return-to-home commands), reliable delivery is required to ensure safe operation. Conversely, high-rate topics (e.g., live video streams) are better suited to best effort delivery, as freshness is prioritized over completeness. Finally, some topics convey information that is important but not safety-critical (e.g., telemetry). As depicted in Fig. \ref{fig_rel}, these topics are candidates for dynamic adaptation, using reliable delivery under favorable conditions and falling back to best effort under degraded conditions.

SDN provides global visibility of the underlying IoD network (e.g., topology, link status, congestion, etc.) and enables dynamic, network-wide control \cite{zormatiReviewAnalysisRecent2024}. We enable global network monitoring by adopting a centralized SDN architecture in which the controller is part of the Ground Control Infrastructure (GCI). Each drone operates as an SDN-enabled forwarding node, capable of receiving flow rules from the controller and reporting local link metrics. We assume an out-of-band control plane, where control traffic between drones and the GCI uses a dedicated wireless link \cite{singhOverviewDroneCommunication2022}.

\begin{figure}[ht]
    \centering
    \includegraphics[scale=0.23]{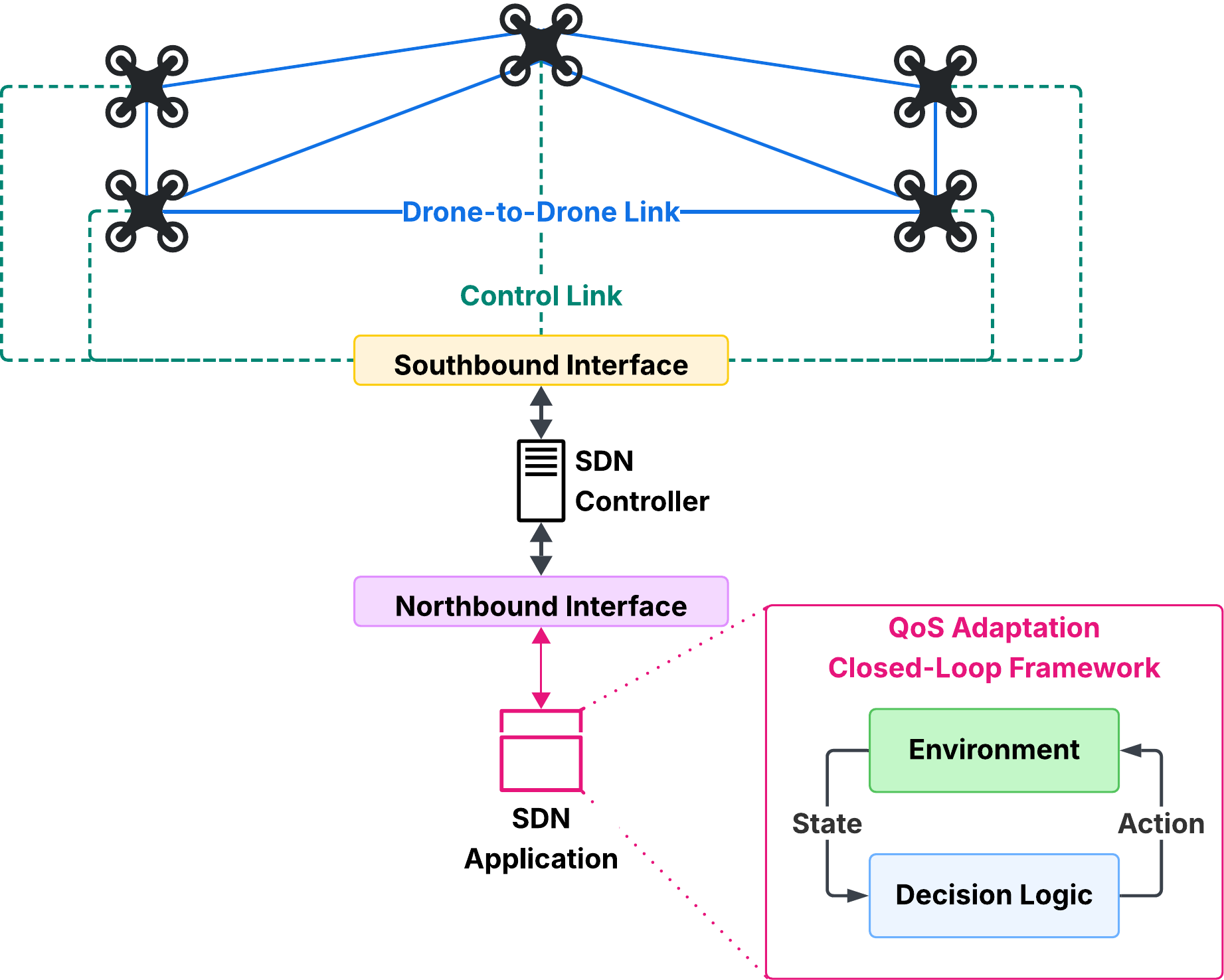}
    \caption{Proposed QoS adaptation framework}
    \label{fig_prop_sol}
\end{figure}

We propose a cross-layer, SDN-assisted, reactive QoS adaptation framework for DDS-based IoD communications that improves communication reliability and efficiency by dynamically adjusting the reliability modes of adaptive topics based on network-level conditions and device-level constraints (e.g., available energy). The proposed model also accounts for topic popularity, as it directly impacts network load and device energy consumption. The adaptation is performed at the topic level as it captures topic-specific characteristics such as popularity that cannot be expressed at the link level. The proposed solution is depicted in Fig. \ref{fig_prop_sol}. The framework is implemented as an SDN application that interacts with the controller through the northbound Application Programming Interface (API). Actions are computed by the QoS adaptation framework and enforced through DDS QoS agents on each UAV under the coordination of the SDN controller.

\section{Proposed Adaptation Solution}
The proposed QoS adaptation framework monitors the operating environment and collects cross-layer system state, including SDN-derived network information. Based on this state, it computes and enforces adaptation decisions at the middleware layer, integrating device-, topic-, and network-level inputs. To formally describe the adaptation process, we define the system state space, the action space, an objective function capturing trade-offs, and the corresponding optimization problem.

\subsection{State space}

The state space captures the network, device, and topic conditions that directly affect the DDS reliability decisions in IoD environments. We define the system state at discrete decision epoch $k$ for a topic $\tau$ as:
\begin{equation}
\mathbf{s}_{k,\tau} = \big[ L_k,\; C_k,\; E_k,\; P_{k,\tau} \big],
\end{equation}
where:

\begin{itemize}
    \item \textbf{Link loss rate ($L_k$):} Packet loss rate observed on active communication links. This network-level metric reflects the reliability of the underlying wireless links.
    \item \textbf{Congestion indicator ($C_k$):} Measure of link utilization (i.e., load). This network-level metric captures the risk of congestion amplification under reliable delivery.
    \item \textbf{Energy level ($E_k$):} Residual battery state of UAVs. This device-level state reflects energy constraints.
    \item \textbf{Topic popularity ($P_{k,\tau}$):} Number of active subscribers to a topic $\tau$. This topic-level metric captures the effect of popularity on the overhead induced by reliable mode.
\end{itemize}

In DDS, dissemination cost depends on both link conditions and the number of subscribers. Under reliable mode, acknowledgments and retransmissions introduce overhead that scales with subscriber count, an effect amplified in swarm-based IoD scenarios \cite{leeOptimizingDataDistribution2025}.

The instantaneous popularity of topic $\tau$ at decision epoch $k$ is defined as:
\begin{equation}
N_{k,\tau} = \left| \mathcal{R}_{k,\tau} \right|,
\end{equation}
where $\mathcal{R}_{k,\tau}$ denotes the set of subscribers (i.e., DDS data readers) currently matched to topic $\tau$. To ensure comparability across different fleet sizes, popularity is normalized \cite{chaudharyPeNCachePopularityBased2025} as:
\begin{equation}
P_{k,\tau} = \frac{N_{k,\tau}}{N_{\max}},
\qquad P_{k,\tau} \in [0,1],
\end{equation}
where $N_{\max}$ denotes the maximum expected number of subscribers (e.g., maximum swarm size).

To reduce sensitivity to transient membership fluctuations, a smoothed popularity estimate is used:
\begin{equation}
\bar{P}_{k,\tau} = (1-\lambda)\bar{P}_{k-1,\tau} + \lambda P_{k,\tau},
\qquad \lambda \in (0,1],
\end{equation}
where $\lambda$ is the smoothing factor.

In the remainder of this work, $P_{k,\tau}$ denotes the smoothed topic popularity estimate $\bar{P}_{k,\tau}$. The metric $P_{k,\tau}$ can be computed using DDS discovery and endpoint matching information. The network metrics $L_k$ and $C_k$ are path-level aggregates obtained from the SDN controller, while $E_k$ represents the residual energy state of the UAVs.

\subsection{Action space}
The action space represents the set of runtime middleware-level adaptations that the QoS adaptation framework can enforce. For clarity, while maintaining DDS standard compliance \cite{OMG_DDS_1_4}, we focus on DDS reliability mode applied to adaptive topics. At each decision epoch $k$ for a topic $\tau$, the action is defined as:
\begin{equation}
a_{k,\tau} = R_{k,\tau}, \quad R_{k,\tau} \in \{\texttt{RELIABLE},\; \texttt{BEST\_EFFORT}\}.
\end{equation}

The \texttt{RELIABLE} mode ensures delivery at the cost of higher latency, bandwidth, and energy, whereas \texttt{BEST\_EFFORT} favors low latency and overhead without guarantees. Only adaptive topics are subject to switching, while safety-critical topics retain fixed QoS settings. Actions are enforced through DDS QoS agents on each UAV.

\subsection{Objective function}
The objective is to select, for each adaptive topic $\tau$, the DDS reliability mode that balances communication performance against reliability-induced overhead under the current operating conditions. Reliability decisions are taken per topic, while critical topics are not subject to adaptation.

To compare heterogeneous metrics within a unified decision framework, all state variables are normalized to the interval $[0,1]$, yielding $\tilde{L}_k$, $\tilde{C}_k$, $\tilde{E}_k$, and $\tilde{P}_{k,\tau}$. Each state variable is normalized to the interval $[0,1]$ using min--max normalization with fixed operational bounds \cite{huangFederatedLearningBased2023}. 

Specifically, for a generic metric $x_k$, normalization is performed as
\begin{equation}
\tilde{x}_k = \min\!\left(1,\; \max\!\left(0,\; \frac{x_k - x_{\min}}{x_{\max} - x_{\min}} \right)\right),
\end{equation}
where $x_{\min}$ and $x_{\max}$ denote the minimum and maximum expected operational values of $x_k$.

The instantaneous cost for topic $\tau$ represents the penalty incurred by choosing a given reliability mode for topic $\tau$ at epoch $k$ and is defined as: 
\begin{equation}
\begin{aligned}
J(\mathbf{s}_{k,\tau}, R_{k,\tau})
={}& w_L\,\tilde{L}_k\,\mathbb{I}\!\left[R_{k,\tau}=\texttt{BEST\_EFFORT}\right] \\
&+ \mathbb{I}\!\left[R_{k,\tau}=\texttt{RELIABLE}\right]
\Big(
w_C\,\tilde{C}_k
+ w_P\,\tilde{P}_{k,\tau} \\
&\hspace{4.2em}
+ w_E\,(1-\tilde{E}_k)
\Big) \\
&+ \mathbb{I}\!\left[R_{k,\tau}=\texttt{RELIABLE}\right]
\Big(
w_{CL}\,\tilde{C}_k\tilde{L}_k \\
&\hspace{4.2em}
+ w_{PL}\,\tilde{P}_{k,\tau}\tilde{L}_k
\Big) \\
&+ w_S\,\mathbb{I}\!\left[R_{k,\tau}\neq R_{k-1,\tau}\right],
\end{aligned}
\end{equation}
where $\mathbb{I}[\cdot]$ denotes the indicator function. The term $w_L \tilde{L}_k$ discourages operation in best effort mode when the loss rate is high. When the reliable mode is selected, additional costs are incurred to capture reliability-induced overhead. Specifically, $w_C \tilde{C}_k$ represents congestion cost due to retransmissions and protocol overhead, $w_P \tilde{P}_{k,\tau}$ penalizes highly popular topics since reliability overhead scales with the number of subscribers, and $w_E (1-\tilde{E}_k)$ discourages the use on energy-constrained UAVs. We further account for loss-amplified reliability overhead (i.e., retransmission storm risk). The term $w_{CL}\,\tilde{C}_k \tilde{L}_k$ captures the coupling between congestion and loss, where packet loss triggers retransmissions while congestion limits available capacity, potentially leading to retransmission storms. Similarly, $w_{PL}\,\tilde{P}_{k,\tau} \tilde{L}_k$ captures the fact that, for popular topics, recovery traffic scales jointly with loss and the number of subscribers. Finally, $w_S$ penalizes frequent switches between reliability modes, mitigating oscillatory behavior.

By tuning the weighting coefficients, we can control the tradeoff between performance and efficiency: larger $w_L$ promotes conservative, reliability-oriented behavior, whereas larger $\{w_C, w_P, w_E\}$ favor efficiency-aware policies that select best effort delivery under constrained conditions.

Given the defined state space, action space, and cost function, the QoS adaptation problem is to select, for each adaptive topic $\tau$ and at each decision epoch $k$, the DDS reliability mode that minimizes the incurred cost. The optimal decision can be obtained via myopic optimization. Specifically, the optimal reliability mode for topic $\tau$ at epoch $k$ is given by:
\begin{equation}
R_{k,\tau}^\star
=
\arg\min_{R_{k,\tau} \in \{\texttt{RELIABLE},\; \texttt{BEST\_EFFORT}\}}
J(\mathbf{s}_{k,\tau}, R_{k,\tau}).
\end{equation}

A myopic formulation is appropriate since reliability decisions primarily affect instantaneous overhead and loss recovery costs, while longer-term coupling across decision epochs is limited and captured via the switching penalty.

\subsection{Myopic solution}
Since the action space is binary, the per-topic optimization admits a closed-form myopic solution obtained by direct comparison of the two candidate actions. Let
$J^{\texttt{BE}}_{k,\tau} = J(\mathbf{s}_{k,\tau},\texttt{BEST\_EFFORT})$
and
$J^{\texttt{REL}}_{k,\tau} = J(\mathbf{s}_{k,\tau},\texttt{RELIABLE})$.
The optimal decision is given by:
\begin{equation}
R_{k,\tau}^\star =
\begin{cases}
\texttt{RELIABLE}, & \text{if } J^{\texttt{REL}}_{k,\tau} < J^{\texttt{BE}}_{k,\tau}, \\
\texttt{BEST\_EFFORT}, & \text{otherwise}.
\end{cases}
\end{equation}

Equivalently, defining the cost difference
\begin{equation}
\label{eq:delta_cost}
\Delta_{k,\tau} = J^{\texttt{REL}}_{k,\tau} - J^{\texttt{BE}}_{k,\tau},
\end{equation}
the controller selects the reliable mode whenever the reliability-induced overhead is outweighed by the loss penalty incurred under best effort operation, i.e., when $\Delta_{k,\tau} < 0$.

Substituting the cost function, the difference between the two actions can be written explicitly as
\begin{equation}
\label{eq:delta_final}
\begin{aligned}
\Delta_{k,\tau}
=\;& J^{\texttt{REL}}_{k,\tau} - J^{\texttt{BE}}_{k,\tau} \\
=\;&
\underbrace{
w_C\,\tilde{C}_k
+ w_P\,\tilde{P}_{k,\tau}
+ w_E\,(1-\tilde{E}_k)
}_{\text{baseline reliability overhead}}
\\
&+
\underbrace{
w_{CL}\,\tilde{C}_k \tilde{L}_k
+ w_{PL}\,\tilde{P}_{k,\tau} \tilde{L}_k
}_{\text{loss-amplified overhead}}
\\
&-
\underbrace{
w_L\,\tilde{L}_k
}_{\text{loss penalty under best effort}}
\\
&+
\underbrace{
w_S\!\Big(
\mathbb{I}\!\left[R_{k-1,\tau}\neq \texttt{RELIABLE}\right]
\Big)
}_{\text{switching bias}}
\\
&-
\underbrace{
w_S\!\Big(
\mathbb{I}\!\left[R_{k-1,\tau}\neq \texttt{BEST\_EFFORT}\right]
\Big)
}_{\text{switching bias}} .
\end{aligned}
\end{equation}

Algorithm \ref{alg:dds_reliability_delta} summarizes  the proposed myopic reliability adaptation procedure. The computational complexity per decision epoch is $\mathcal{O}(|\mathcal{T}_A|)$, where $|\mathcal{T}_A|$ denotes the number of adaptive topics. The required memory scales as $\mathcal{O}(1)$ per topic.

\section{Experimental Results}
This section evaluates the proposed framework through simulation. We first present the experimental setup. Then, we present the simulation results and discuss the findings. 

\subsection{Simulation environment}
We conduct experiments using Mininet, a network emulator, with an SDN control plane implemented via the Ryu controller and Cyclone DDS as middleware. We consider small-scale UAV communication topologies with background traffic generation. Network conditions are varied through controlled link impairments and bandwidth constraints, while device-level constraints are captured via a software-based energy model. This setup enables the evaluation of the framework under representative wireless IoD communication dynamics.

\begin{figure}[!t]

\begin{minipage}{\columnwidth}
\begin{algorithm}[H]
\caption{Myopic DDS Reliability Adaptation}
\label{alg:dds_reliability_delta}
\begin{algorithmic}[0]
\State \textbf{Input:} Weights $\{w_L, w_C, w_P, w_E, w_{CL}, w_{PL}, w_S\}$
\State \textbf{Output:} Reliability mode $R_{k,\tau}^\star$ for adaptive topics

\For{each decision epoch $k$}
    \State Observe normalized state $\{\tilde{L}_k, \tilde{C}_k, \tilde{E}_k, \tilde{P}_{k,\tau}\}$
    \ForAll{adaptive topics $\tau$}
        \State Compute $\Delta_{k,\tau}$ using~\eqref{eq:delta_final}
        \If{$\Delta_{k,\tau} < 0$}
            \State $R_{k,\tau}^\star \gets \texttt{RELIABLE}$
        \Else
            \State $R_{k,\tau}^\star \gets \texttt{BEST\_EFFORT}$
        \EndIf
        \State Enforce $R_{k,\tau}^\star$ via DDS QoS agent
    \EndFor
\EndFor
\end{algorithmic}
\end{algorithm}
\end{minipage}
\end{figure}

\subsection{Results and discussion}
We conduct a set of simulations to evaluate the effectiveness of our solution. We focus on representative scenarios to isolate and validate each adaptation component. This allows us to assess the behavior of the framework under controlled network impairments and compare its performance to that of static configurations. In what follows, we progressively activate the decision components of the framework. Experiments are repeated over 20 independent runs for each parameter setting (e.g., loss or congestion level), each lasting 100 seconds (including warm-up and cool-down) to ensure steady-state conditions. The results are aggregated after extreme transient deviations are excluded using the Tukey Inter-Quartile Range (IQR) criterion.

\subsubsection{Message delivery under lossy links}
\begin{figure}[ht]
    \centering
    \includegraphics[scale=0.33]{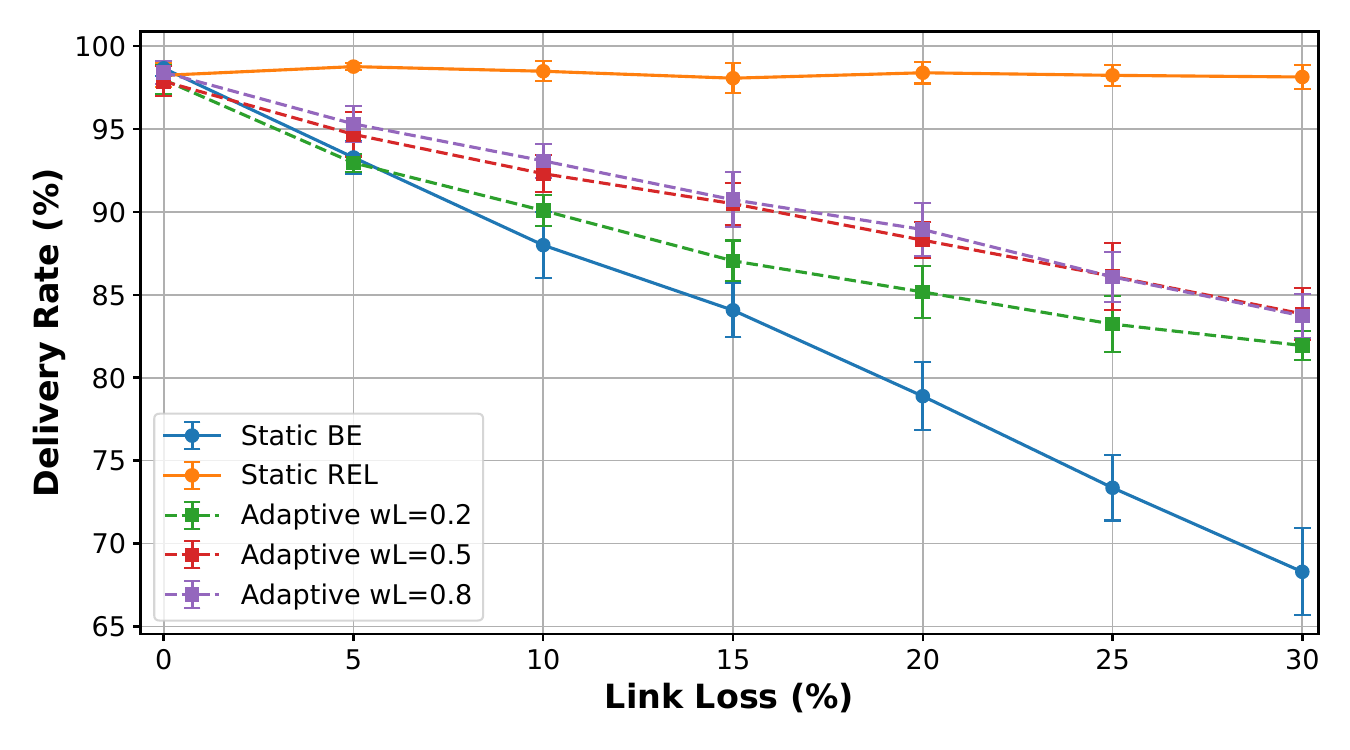}
    \caption{Message delivery rate vs. link loss}
    \label{fig_res_dr_loss}
\end{figure}

~\\
\indent We measure the message delivery rate (i.e., the ratio of received messages to transmitted messages) under increasing link loss (0–30\%) and compare three configurations: static best effort (BE), static reliable (REL), and adaptive reliability for $w_L \in \{0.2, 0.5, 0.8\}$, while setting $w_C = 0$ and $w_E = 0$. Results are depicted in Fig. \ref{fig_res_dr_loss}.

Static best effort shows a progressive degradation in delivery rate as link loss increases due to the absence of retransmissions, whereas static reliable maintains high delivery performance across all loss levels at the cost of additional overhead. The adaptive configuration dynamically switches between best effort and reliable modes according to the observed state. The weight parameter $w_L$ controls the emphasis on reliability: lower values favor best effort, while higher values favor reliable transmission. Under low-loss conditions, the framework predominantly selects best effort to preserve efficiency. As loss increases, it progressively shifts toward reliable mode, improving delivery performance. Increasing $w_L$ moves the adaptive behavior closer to static reliable operation, enabling a tunable reliability--efficiency trade-off.

Based on these results, we conclude that the proposed adaptive framework responds consistently to varying loss conditions while maintaining performance.

\subsubsection{Communication overhead under congestion}
\begin{figure}[ht]
    \centering
    \includegraphics[scale=0.35]{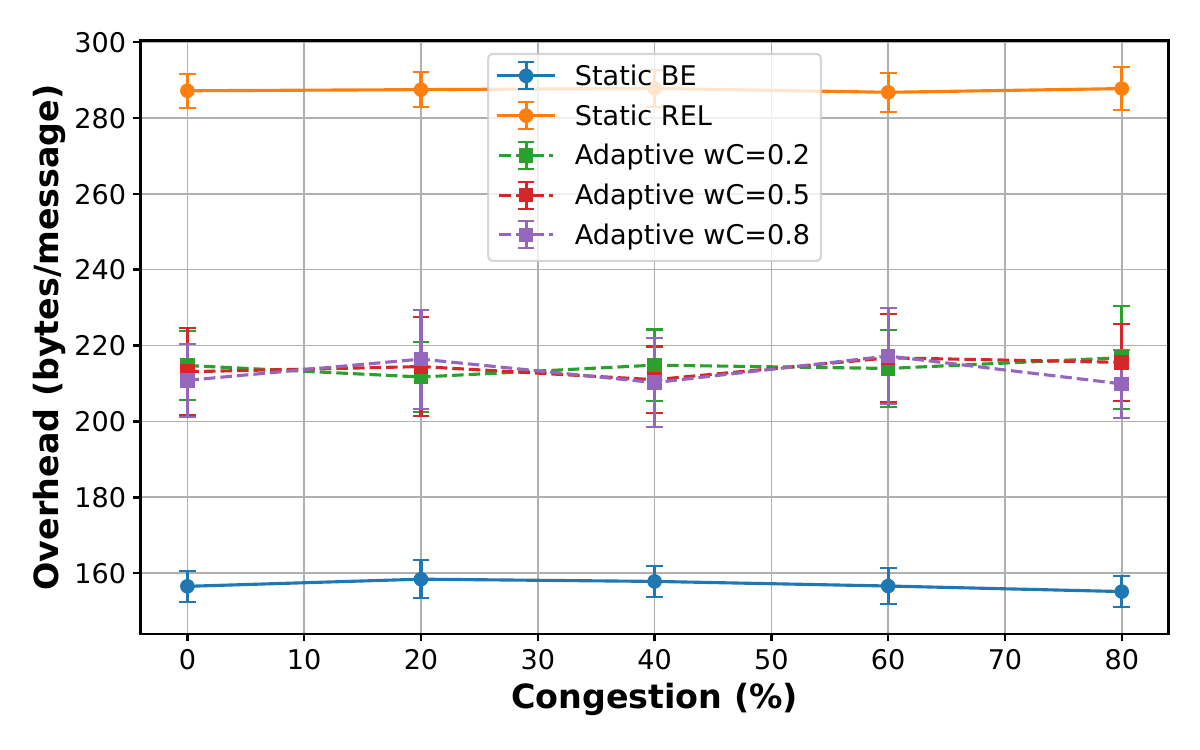}
    \caption{Communication overhead vs. congestion}
    \label{fig_res_overhead_cong}
\end{figure}

~\\
\indent We evaluate the communication overhead under increasing network congestion (0--80\%). Congestion is emulated via background traffic consuming a controlled share of the bottleneck capacity. Overhead is defined as the total transmitted bytes divided by successfully delivered messages. We compare static best effort, static reliable, and adaptive configurations for $w_C \in \{0.2, 0.5, 0.8\}$ while keeping $w_L = 0.5$ and $w_E = 0$. Results are shown in Fig. \ref{fig_res_overhead_cong}.

Static reliable exhibits consistently higher overhead across all congestion levels due to acknowledgment traffic and retransmissions, whereas static best effort maintains the lowest overhead by avoiding retransmission. The adaptive configuration adjusts the reliability mode according to the congestion indicator. Under low congestion, it behaves similarly to static reliable to preserve delivery robustness. As congestion increases, it progressively favors best effort mode, reducing retransmission overhead and limiting additional load on saturated links. Increasing $w_C$ strengthens this congestion-awareness, shifting the adaptive behavior closer to static best effort and enabling a tunable reliability--efficiency trade-off.

These findings validate the congestion-awareness component of the framework and demonstrate its ability to reduce network load in constrained conditions.

\subsubsection{Energy efficiency under battery constraints}
\begin{figure}[ht]
    \centering
    \includegraphics[scale=0.274]{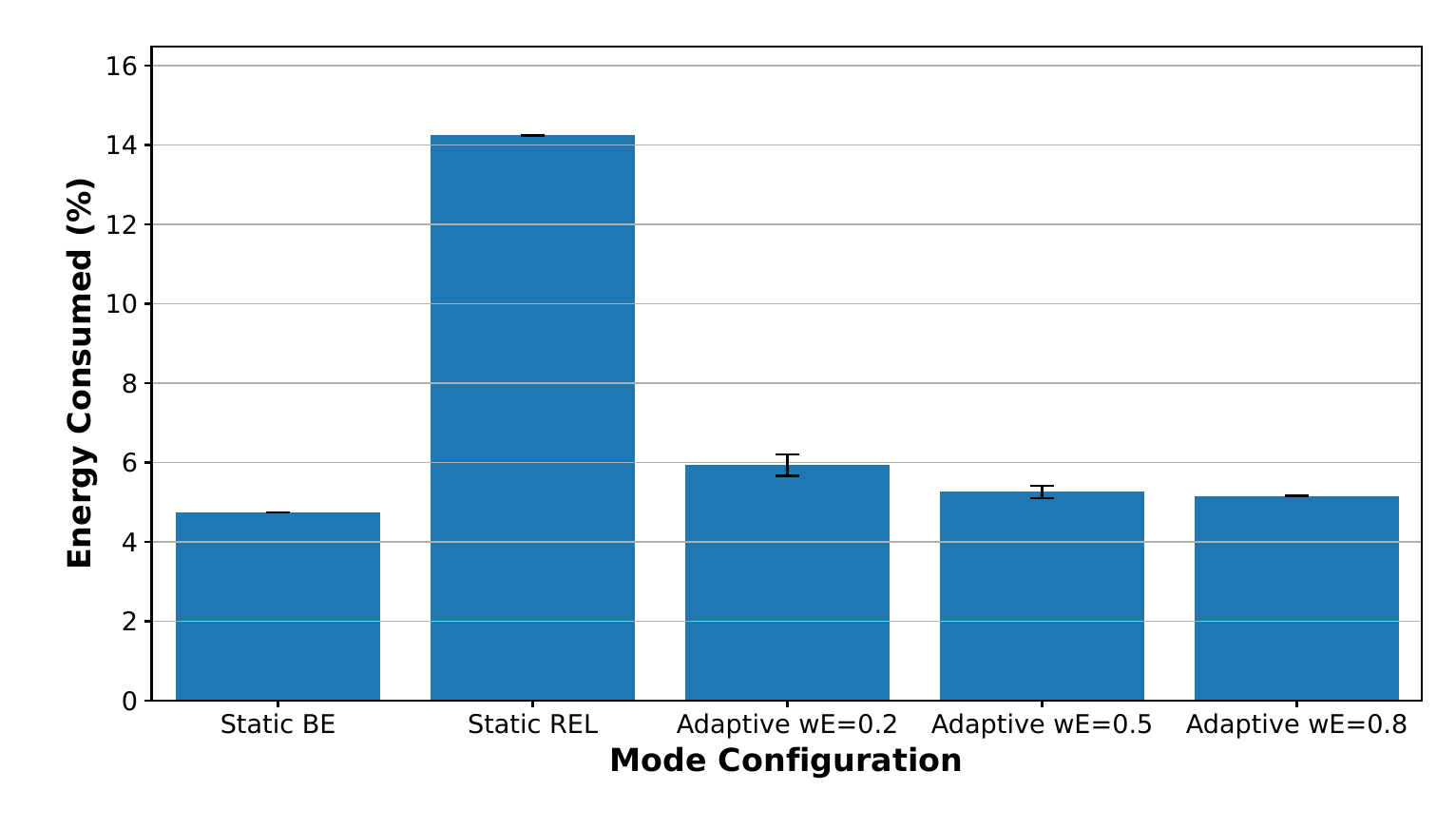}
    \caption{Comparison of energy consumption}
    \label{fig_res_energy_comp}
\end{figure}

~\\
\indent We finally evaluate the energy-awareness component of the framework. In this experiment, we introduce a battery state variable that decreases proportionally to transmission activity. We compare static best effort, static reliable, and the adaptive configuration for $w_E \in \{0.2, 0.5, 0.8\}$, while keeping $w_L = 0.5$ and $w_C = 0.5$. Energy consumption is computed as the normalized battery depletion during each run. Results are shown in Fig. \ref{fig_res_energy_comp}.

Static reliable exhibits the highest energy consumption due to acknowledgment traffic and retransmissions, whereas static best effort minimizes energy usage by avoiding retransmission overhead. The adaptive configuration adjusts the reliability mode based on the battery state through the weight $w_E$. When energy levels are high, reliable transmission is favored to maintain robustness. As the battery depletes, the framework progressively shifts toward best effort to reduce transmission cost and extend operational lifetime. Increasing $w_E$ amplifies energy awareness, moving the adaptive behavior closer to static best effort and lowering overall energy consumption.

The results show that the framework effectively incorporates energy awareness, offering a controllable trade-off between reliability and resource efficiency to prolong operation under limited battery conditions.

Across all scenarios, the framework shows consistent, controllable behavior under varying constraints: it increases reliable transmission over lossy links, shifts to best effort under congestion to reduce overhead, and favors energy-efficient operation under battery limits to extend system lifetime. These results demonstrate effective integration of reliability, congestion-awareness, and energy-awareness into a unified, context-aware DDS adaptation mechanism.

\section{Conclusion}
In this work, we proposed a cross-layer DDS reliability adaptation framework that dynamically adjusts communication reliability by jointly considering global network conditions, device-level energy constraints, and topic popularity. Simulation results demonstrate the effectiveness of the proposed framework in improving communication reliability and efficiency in IoD scenarios.

As future work, we plan to extend the framework to incorporate additional DDS QoS policies (e.g., deadline and history depth), investigate proactive adaptation strategies that anticipate network degradation and mobility-induced disruptions, and explore learning-based approaches for adaptive tuning of decision parameters. We also aim to validate the framework through extensive overhead and sensitivity analyses and through experimental deployment on the Heudiasyc drone testbed.

\bibliographystyle{IEEEtran}
\bibliography{IEEEbib}

@article{abualigahApplicationsDeploymentsIntegration2021,
  author  = {Abualigah, Laith and Diabat, Ali and Sumari, Putra and Gandomi, Amir H.},
  title   = {Applications, Deployments, and Integration of Internet of Drones ({IoD}): A Review},
  journal = {IEEE Sensors Journal},
  volume  = {21},
  number  = {22},
  pages   = {25532--25546},
  year    = {2021},
  doi     = {10.1109/JSEN.2021.3114266}
}

@article{bineConnectingInternetDrones2024,
  author  = {Bine, Lailla M. S. and Boukerche, Azzedine and Ruiz, Linnyer B. and Loureiro, Antonio A. F.},
  title   = {Connecting Internet of Drones and Urban Computing: Methods, Protocols and Applications},
  journal = {Computer Networks},
  volume  = {239},
  pages   = {110136},
  year    = {2024},
  doi     = {10.1016/j.comnet.2023.110136}
}

@article{boccadoroExtensiveSurveyInternet2021,
  author  = {Boccadoro, Pietro and Striccoli, Domenico and Grieco, Luigi Alfredo},
  title   = {An Extensive Survey on the Internet of Drones},
  journal = {Ad Hoc Networks},
  volume  = {122},
  pages   = {102600},
  year    = {2021},
  doi     = {10.1016/j.adhoc.2021.102600}
}

@article{chaudharyPeNCachePopularityBased2025,
  author  = {Chaudhary, Pankaj and Hubballi, Neminath},
  title   = {{PeNCache}: Popularity Based Cooperative Caching in Named Data Networks},
  journal = {Computer Networks},
  volume  = {257},
  pages   = {110995},
  year    = {2025},
  doi     = {10.1016/j.comnet.2024.110995}
}

@techreport{OMG_DDS_1_4,
  author      = {{Object Management Group}},
  title       = {Data Distribution Service ({DDS}) Version 1.4},
  institution = {OMG},
  year        = {2015},
  number      = {formal/2015-04-10}
}

@article{huangFederatedLearningBased2023,
  author  = {Huang, Xiaoge and Chen, Zhi and Chen, Qianbin and Zhang, Jie},
  title   = {Federated Learning Based {QoS}-Aware Caching Decisions in Fog-Enabled Internet of Things Networks},
  journal = {Digital Communications and Networks},
  volume  = {9},
  number  = {2},
  pages   = {580--589},
  year    = {2023},
  doi     = {10.1016/j.dcan.2022.04.022}
}

@article{javedStateoftheArtFutureResearch2024,
  author  = {Javed, Sadaf and Hassan, Ali and Ahmad, Rizwan and Ahmed, Waqas and Ahmed, Rehan and Saadat, Ahsan and Guizani, Mohsen},
  title   = {State-of-the-Art and Future Research Challenges in {UAV} Swarms},
  journal = {IEEE Internet of Things Journal},
  volume  = {11},
  number  = {11},
  pages   = {19023--19045},
  year    = {2024},
  doi     = {10.1109/JIOT.2024.3364230}
}

@article{zormatiReviewAnalysisRecent2024,
  author  = {Zormati, Mohamed Ali and Lakhlef, Hicham and Ouni, Sofiane},
  title   = {Review and Analysis of Recent Advances in Intelligent Network Softwarization for the Internet of Things},
  journal = {Computer Networks},
  volume  = {241},
  pages   = {110215},
  year    = {2024},
  doi     = {10.1016/j.comnet.2024.110215}
}

@article{leeOptimizingDataDistribution2025,
  author  = {Lee, HyeonGyu and Kim, Doyoon and Moon, SungTae},
  title   = {Optimizing Data Distribution Service Discovery for Swarm {UAVs} Through Preloading and Network Awareness},
  journal = {Drones},
  volume  = {9},
  number  = {8},
  year    = {2025},
  doi     = {10.3390/drones9080564}
}

@article{gamboSystematicLiteratureReview2025,
  author  = {Gambo, Muhammad Liman and Danasabe, Abubakar and Almadani, Basem and Aliyu, Farouq and Aliyu, Abdulrahman and Al-Nahari, Esam},
  title   = {A Systematic Literature Review of {DDS} Middleware in Robotic Systems},
  journal = {Robotics},
  volume  = {14},
  number  = {5},
  year    = {2025},
  doi     = {10.3390/robotics14050063}
}

@article{zhangComparisonMiddlewaresEdgetoEdge2024,
  author  = {Zhang, Jiaqiang and Yu, Xianjia and Ha, Sier and Peña Queralta, Jorge and Westerlund, Tomi},
  title   = {Comparison of Middlewares in Edge-to-Edge and Edge-to-Cloud Communication for Distributed {ROS} 2 Systems},
  journal = {Journal of Intelligent \& Robotic Systems},
  volume  = {110},
  number  = {4},
  pages   = {162},
  year    = {2024},
  doi     = {10.1007/s10846-024-02187-z}
}

@article{alaerjanFormalizingSemanticsDDS2023,
  author  = {Alaerjan, Alaa},
  title   = {Formalizing the Semantics of {DDS} {QoS} Policies for Improved Communications in Distributed Smart Grid Applications},
  journal = {Electronics},
  volume  = {12},
  number  = {10},
  year    = {2023},
  doi     = {10.3390/electronics12102246}
}

@article{ingles-romeroModelDrivenApproachEnable2017,
  author  = {Inglés-Romero, Juan F. and Romero-Garcés, Adrián and Vicente-Chicote, Cristina and Martínez, Jesús},
  title   = {A Model-Driven Approach to Enable Adaptive {QoS} in {DDS}-Based Middleware},
  journal = {IEEE Trans. Emerging Topics Comput. Intell.},
  volume  = {1},
  number  = {3},
  pages   = {176--187},
  year    = {2017},
  doi     = {10.1109/TETCI.2017.2669187}
}

@inproceedings{yoonQoSOptimizer2016,
  author    = {Yoon, Gunjae and Lee, Shineun and Choi, Hoon},
  title     = {{QoS} Optimizer},
  booktitle = {Proc. Int. Conf. Platform Technology and Service (PlatCon)},
  pages     = {1--5},
  year      = {2016},
  doi       = {10.1109/PlatCon.2016.7456819}
}

@inproceedings{bertauxDDSSDNBased2014a,
  author    = {Bertaux, Lionel and Hakiri, Akram and Medjiah, Samir and Berthou, Pascal and Abdellatif, Slim},
  title     = {A {DDS}/{SDN}-Based Communication System for Efficient Support of Dynamic Distributed Real-Time Applications},
  booktitle = {Proc. IEEE/ACM Int. Symp. Distributed Simulation and Real-Time Applications},
  pages     = {77--84},
  year      = {2014},
  doi       = {10.1109/DS-RT.2014.18}
}

@inproceedings{choiMakingDDSReally2016,
  author    = {Choi, Hyon-Young and King, Andrew L. and Lee, Insup},
  title     = {Making {DDS} Really Real-Time with {OpenFlow}},
  booktitle = {Proc. Int. Conf. Embedded Software (EMSOFT)},
  pages     = {1--10},
  year      = {2016},
  doi       = {10.1145/2968478.2968479}
}

@article{liEnablingDeterministicTransmission2025,
  author  = {Li, Binqi and Zhu, Yuan and Yao, Xiangxi and Jiang, Chenming and Lu, Ke and Sun, Zhipeng},
  title   = {Enabling Deterministic Transmission for {DDS} by Leveraging {IEEE} 802.1{Qbv} Time-Sensitive Networking},
  journal = {Computer Networks},
  volume  = {261},
  pages   = {111128},
  year    = {2025},
  doi     = {10.1016/j.comnet.2025.111128}
}

@article{al-roubaieyEATDDSEnergyawareMiddleware2019,
  author  = {Al-Roubaiey, Anas and Sheltami, Tarek and Mahmoud, Ashraf and Yasar, Ansar},
  title   = {{EATDDS}: Energy-Aware Middleware for Wireless Sensor and Actuator Networks},
  journal = {Future Generation Computer Systems},
  volume  = {96},
  pages   = {196--206},
  year    = {2019},
  doi     = {10.1016/j.future.2019.01.060}
}

@inproceedings{salehiPopSubImprovingResource2017,
  author    = {Salehi, Pooya and Zhang, Kaiwen and Jacobsen, Hans-Arno},
  title     = {{PopSub}: Improving Resource Utilization in Distributed Content-Based Publish/Subscribe Systems},
  booktitle = {Proc. ACM Int. Conf. Distributed Event-Based Systems (DEBS)},
  pages     = {88--99},
  year      = {2017},
  doi       = {10.1145/3093742.3093915}
}

@article{vidalEnablingMultiMissionInteroperable2018,
  author  = {Vidal, Ivan and Bellavista, Paolo and Sanchez-Aguero, Victor and Garcia-Reinoso, Jaime and Valera, Francisco and Nogales, Borja and Azcorra, Arturo},
  title   = {Enabling Multi-Mission Interoperable {UAS} Using Data-Centric Communications},
  journal = {Sensors},
  volume  = {18},
  number  = {10},
  year    = {2018},
  doi     = {10.3390/s18103421}
}

@inproceedings{singhOverviewDroneCommunication2022,
  author = {R. Singh and K. D. Ballal and M. S. Berger and L. Dittmann},
  title = {Overview of Drone Communication Requirements in {5G}},
  booktitle = {Proc. IoT Week},
  pages = {3--16},
  year = {2022},
  doi = {10.1007/978-3-031-20936-9_1}
}

\end{document}